\documentclass[11pt,a4paper]{article}
\pdfoutput=1
\usepackage{jheppub}

\usepackage{ifpdf}
\usepackage{afterpage}
\usepackage{amssymb}
\usepackage{amsmath}
\usepackage{graphicx}
\usepackage{longtable}
\usepackage{verbatim}
\usepackage{amsfonts}
\usepackage{bbold}
\usepackage[table,usenames,dvipsnames]{xcolor}
\usepackage{titlesec}
\usepackage{enumerate}

\usepackage{tikz}
\usetikzlibrary{trees}
\usetikzlibrary{decorations.pathmorphing}
\usetikzlibrary{decorations.markings}
\tikzset{
    photon/.style={decorate, decoration={snake}, draw=black},
    wino/.style={draw=redwine},
    electron/.style={draw=black, postaction={decorate},
        decoration={markings,mark=at position .55 with {\arrow[draw=black]{>}}}},
    scalar/.style={draw=black, dashed,postaction={decorate},
        decoration={markings,mark=at position .55 with {\arrow[draw=black]{>}}}},
    gluon/.style={decorate, draw=black,
        decoration={coil,amplitude=4pt, segment length=5pt}}
}

\newcommand{\bear}{\begin{array}}
\newcommand{\ear}{\end{array}}
\newcommand{\beq}{\begin{equation}}
\newcommand{\eeq}{\end{equation}}
\newcommand{\beqa}{\begin{eqnarray}}
\newcommand{\eeqa}{\end{eqnarray}}

\def\OMIT#1{{}}
\newcommand{\lsim}{\mathrel{\rlap{\lower4pt\hbox{\hskip1pt$\sim$}}
    \raise1pt\hbox{$<$}}}         
\newcommand{\gsim}{\mathrel{\rlap{\lower4pt\hbox{\hskip1pt$\sim$}}
    \raise1pt\hbox{$>$}}}         

\newcommand{\bl}{\left}
\newcommand{\br}{\right}

\newcommand{\ie}{\textit{i.e.}\ }

\newcommand{\dd}{\mathrm{d}}

\newcommand{\mL}{\mathcal{L}}
\newcommand{\mO}{\mathcal{O}}

\newcommand{\tot}{\mathrm{tot}}

\newcommand{\vir}{\mathrm{vir}}

\newcommand{\igm}{\mathrm{IGM}}
\newcommand{\icm}{\mathrm{ICM}}

\newcommand{\gam}{\gamma}
\newcommand{\ag}{{a\gamma}}
\newcommand{\reff}{\mathrm{ref}}
\newcommand{\ini}{\mathrm{ini}}
\newcommand{\eff}{\mathrm{eff}}

\newcommand{\maxim}{\mathrm{max}}
\newcommand{\Pan}{\mathrm{Pan}}
\newcommand{\cl}{\mathrm{cl}}

\newcommand{\BAO}{\mathrm{BAO}}
\newcommand{\SHOES}{\mathrm{SH0ES}}
\newcommand{\TDCOSMO}{\mathrm{TD}}

\newcommand{\bfth}{\boldsymbol{\theta}}
\newcommand{\bfTH}{\boldsymbol{\Theta}}
\newcommand{\bfxi}{\boldsymbol{\xi}}
\newcommand{\bfa}{\boldsymbol{\alpha}}
\newcommand{\Planck}{\mathrm{Pl}}
\newcommand{\early}{\mathrm{early}}
\newcommand{\late}{\mathrm{late}}

\newcommand{\LC}{\Lambda\mathrm{CDM}}
\newcommand{\rsd}{r_s^\mathrm{drag}}
\newcommand{\gag}{g_{a\gamma\gamma}}
\newcommand{\Pag}{P_{a\gamma}}
\newcommand{\Pgg}{P_{\gamma\gamma}}
\newcommand{\avgPggicm}{\langle \Pgg^\icm \rangle}
\newcommand{\avgPggicmpars}{\langle \Pgg^\icm (m_a, g_{a\gamma\gamma}) \rangle}

\newcommand{\eV}{\mathrm{eV}}

\newcommand{\GeV}{\mathrm{GeV}}

\newcommand{\cm}{\mathrm{cm}}

\newcommand{\km}{\mathrm{km}}
\newcommand{\kpc}{\mathrm{kpc}}
\newcommand{\Mpc}{\mathrm{Mpc}}

\newcommand{\Ga}{\mathrm{G}}
\newcommand{\nGa}{\mathrm{nG}}
\newcommand{\muGa}{\mu\mathrm{G}}

\newcommand{\Sec}[1]{Sec.~\ref{#1}}

\newcommand{\Fig}[1]{Fig.~\ref{#1}}
\newcommand{\Eq}[1]{Eq.~(\ref{#1})}
\newcommand{\Eqs}[2]{Eqs.~(\ref{#1}) and (\ref{#2})}
\newcommand{\Eqsto}[2]{Eqs.~(\ref{#1})-(\ref{#2})}

\newcommand{\ignore}[1]{}

\vspace*{-30mm}

\title{\boldmath Constraints on Axions from Cosmic Distance Measurements}
\author[a]{Manuel A. Buen-Abad,}
\author[a,b]{JiJi Fan,}
\author[c]{Chen Sun}
\affiliation[a]{Department of Physics, Brown University, Providence, RI, 02912, USA}
\affiliation[b]{Brown Theoretical Physics Center, Brown University,
Providence, RI, 02912, U.S.A.}
\affiliation[c]{School of Physics and Astronomy, Tel-Aviv University, Tel-Aviv 69978, Israel}
\emailAdd{manuel\_buen-abad@brown.edu}
\emailAdd{jiji\_fan@brown.edu}
\emailAdd{chensun@mail.tau.ac.il}

\vspace*{1cm}

\abstract{Axion couplings to photons could induce photon-axion conversion in the presence of magnetic fields in the Universe. This conversion could impact various cosmic distance measurements, such as luminosity distances to type Ia supernovae and angular distances to galaxy clusters, in different ways. In this paper we consider different combinations of the most up-to-date distance measurements to constrain the axion-photon coupling. Employing the conservative cell magnetic field model for the magnetic fields in the intergalactic medium (IGM) and ignoring the conversion in the intracluster medium (ICM), we find the upper bounds on axion-photon couplings to be around $5 \times 10^{-12}$ (nG/$B$) $\sqrt{\Mpc/s}$ GeV$^{-1}$ for axion masses $m_a$ below $10^{-13}$ eV, where $B$ is the strength of the IGM magnetic field, and $s$ is the comoving size of the magnetic domains. When including the conversion in the ICM, the upper bound is lowered and could reach $5 \times 10^{-13}\, $GeV$^{-1}$ for $m_a < 5 \times 10^{-12}$ eV. While this stronger bound depends on the ICM modeling, it is independent of the strength of the IGM magnetic field, for which there is no direct evidence yet. These constraints could be placed on firmer footing with an enhanced understanding and control of the astrophysical uncertainties associated with the IGM and ICM. All the bounds are determined by the shape of the Hubble rate as a function of redshift reconstructable from various distance measurements, and insensitive to today's Hubble rate, of which there is a tension between early and late cosmological measurements. As an appendix, we discuss the model building challenges of the use of photon-axion conversion to make type Ia supernovae brighter to alleviate the Hubble problem/crisis.
}

\begin{document}

\maketitle


\section{Introduction}
\label{sec:intro}

Axions, as periodic scalar fields, arise ubiquitously in both low-energy phenomenological models~\cite{Peccei:1977hh, Peccei:1977ur, Weinberg:1977ma, Wilczek:1977pj, Kim:1979if, Shifman:1979if, Zhitnitsky:1980tq, Dine:1981rt} and quantum gravity theories~\cite{Svrcek:2006yi}. They serve as an important benchmark of feebly-coupled light particles beyond the Standard Model (SM). In particular, one of the most active experimental and observational targets is the coupling of an axion, $a$, to photons, which takes the form 
\beq\label{eq:Lag}
    \mL_\ag = - \frac{\gag}{4} a F_{\mu\nu}\tilde{F}^{\mu\nu} = \gag ~ a \mathbf{E}\cdot\mathbf{B} \ ,
\eeq
where $F^{\mu\nu}$ is the electromagnetic field strength, $\tilde{F}^{\mu\nu}=\frac{1}{2} \epsilon^{\mu\nu\rho\sigma}F_{\rho\sigma}$ is its dual field strength, and $\mathbf{E}$ and $\mathbf{B}$ are the electric and magnetic fields. The axion-photon coupling coefficient $\gag$ has mass dimension $-1$ and is inversely proportional to a high energy scale. Various introductions to axion physics basics can be found in ~\cite{Sikivie:2006ni, Marsh:2015xka, Hook:2018dlk, Irastorza:2018dyq}.

On the other hand, the past 20 years have seen an increased interest and corresponding progress in the efforts to measure various cosmic distances to chart out the expansion history of our Universe. One outstanding example is the measurements of luminosity distances (LD), $D_L$, to Type Ia supernovae (SNIa). SNIa are used as ``standard candles" in the Universe given their very similar peak brightnesses. A large number of SN surveys have brought about the Pantheon dataset, which is the largest and most accurate SNIa compilation at present~\cite{Scolnic:2017caz}. It consists of a total of 1048 SNIa in the redshift range of $0.01 <z <2.3$, which can be used to constrain $D_L$ as a function of redshift $z$. Since $D_L(z)$ is determined by the Hubble expansion rate $H(z)$, the Pantheon sample could consequently determine the shape of $H(z)$, at late times. Type Ia supernovae samples are also a crucial input for late-time measurement of today's expansion rate $H_0$, SH0ES~\cite{Riess:2016jrr, Riess:2019cxk}, which is seriously at odds with the early time determination using the CMB data collected by the Planck satellite~\cite{Aghanim:2018eyx}. This is dubbed the ``Hubble problem'' or ``Hubble crisis'' (see \cite{Aylor:2018drw,Knox:2019rjx} and references therein). In addition to LD measurements, there have been several kinds of precise measurements of angular diameter distance (ADD), which is defined as $D_A=d/\theta$ for an astrophysical object of physical size $d$ and angular size $\theta$. Two examples we will use in this paper are from Baryonic Acoustic Oscillations (BAO)~\cite{Beutler:2011hx,Ross:2014qpa,Alam:2016hwk} and galaxy clusters~\cite{DeFilippis:2005hx,Bonamente:2005ct}.

These two seemingly unrelated subjects (axions and their couplings to photons in particle physics on the one hand, and cosmic distances in cosmology on the other) have an intriguing connection. The coupling of axions to photons in Eq.~\eqref{eq:Lag} suggests that in the presence of an external magnetic field photons could convert into axions, and vice versa. Indeed, since there could exist non-negligible magnetic fields in the intergalactic medium (IGM) and/or in the intracluster medium (ICM), the propagation of photons from astrophysical sources could be affected by their conversion into light axions in certain regions of the axion parameter space. This in turn could affect the inference of various distance observables. Take SNIa for example. Assuming the standard model of cosmology, $\Lambda$CDM, photons in the optical band converting into axions could result in a significant dimming of SNIa at higher $z$'s, while SNIa at lower $z$'s are less affected or not affected at all. An effective luminosity distance $D_L(z)$ which takes into account photon-axion conversion could then be constrained by the Pantheon sample, which is consistent with the prediction of pure $\Lambda$CDM. Other distance observables may be modified by photon-axion conversion as well, albeit in different ways. One such example is the angular diameter distance $D_A(z)$ to galaxy clusters. In summary, conversion of photons into axions is effectively equivalent to a departure of the Hubble diagram, $H(z)$, from that of $\Lambda$CDM at late times, which could be constrained by various combinations of cosmic distance measurements.

In this work, we consider different combinations of cosmic distance measurements and carry out statistical analyses to map out the allowed parameter space in the plane of the axion's mass and coupling to photons, $m_a$ and $\gag$ respectively. Analyses using cosmic distance measurements have been performed before, e.g., in Refs.~\cite{Avgoustidis:2010ju, Liao:2015ccl, Tiwari:2016cps} with an older and smaller dataset of SNIa. In addition to including more and updated datasets, our analyses differ from previous ones in the choice of observables. Earlier works usually interpret the constraints as arising from violations of the ``Etherington relation"~\cite{1933PMag}, the distance duality relation between $D_L$ and $D_A$: $D_L(z)= (1+z)^2 D_A(z)$. In other words, the chosen observable is the ratio $D_L/D_A$. It is then (implicitly) assumed that the violation due to photon-axion conversion could be parametrized by a single parameter, e.g., $\epsilon$ such that $D_L = D_A(1+z)^{2+\epsilon}$. As we will discuss in detail, depending on the datasets involved, $D_L$ and $D_A$ could be affected by photon-axion conversion in very different ways and the photon-axion conversion may {\it not} be encoded in a single function of $z$ or a single parameter. Instead, we simply choose the observables to be those quantities directly measured or inferred in each dataset, such as the apparent magnitude of SNIa, the ADDs to galaxy clusters, and the characteristic angular scale of the matter two-point correlation function for BAO; and build corresponding likelihood functions.

The paper is organized as follows: in Sec.~\ref{sec:axion-photon}, we discuss the basic formalism of axion-photon conversion in the IGM or the ICM. We also explain how $D_L$ and $D_A$ could be affected by the conversion. In Sec.~\ref{sec:data}, we describe the datasets included in our analyses and the statistical method we use. In Sec.~\ref{sec:results}, we present and discuss the results as constraints on the axion parameter space. We conclude in Sec.~\ref{sec:concl}. Throughout the paper, we assume that there is negligible axion production at SNIa and that, consequently, the effect of photon-axion conversion in IGM is to dim the SNs. In Appendix~\ref{app:A}, we will entertain the readers with the possibility of resonant axion production at SNIa, which might open up the possibility of brightening SNIa through IGM conversion. We will explain the related model building challenges and why this could not work as a solution to the current Hubble problem/crisis.

\section{Axion-photon conversion}
\label{sec:axion-photon}
In this section, we will first review the basic formulas that describe photon-axion conversion in a magnetic field. We will then discuss the models of the two media, IGM and ICM, in which the conversion takes place. Lastly, we will discuss how various cosmic distances, $D_L$ to SNIa and $D_A$ to galaxy clusters, could be affected by the conversion in different ways. 

Throughout the rest of this paper we will make frequent reference to the parameters that describe axion-photon conversion in a flat $\LC$ cosmological setting. As a shorthand, we denote these parameters as $\bfth = \{ \Omega_\Lambda, H_0, m_a, \gag \}$ and $\bfTH = \bfth \cup \{ M, \rsd \}$; where $\Omega_\Lambda$ is the fraction of today's energy density in the cosmological constant, $H_0$ is the Hubble parameter, $m_a$ is the axion mass, $\gag$ is the axion-photon coupling, $M$ is the absolute magnitude of the SNIa standard candles, and $\rsd$ is the comoving sound horizon size at the time of baryon drag.

\subsection{Basic formulas}
\label{subsec:fund}

In the presence of external magnetic fields, the operator in Eq.~\eqref{eq:Lag} implies that the propagation eigenstates of the photon-axion system are mixtures of axion and photon states. As a result, there is a non-zero probability $P_0$ that a photon oscillates and converts into an axion while traveling through the magnetic field, effectively resulting in photon number violation. When birefringence and Faraday rotation effects are small, as is the case with propagation in the IGM \cite{Mirizzi:2006zy}, the axion mixes only with the photon polarization parallel to the component of the magnetic field $\mathbf{B}_T$, which is transversal to the direction of motion. In the simple case of photons with energy $\omega$ propagating in a constant and homogeneous magnetic field with $B = \vert \mathbf{B}_T \vert$, the axion-photon conversion probability is given by the well-known formula \cite{Georgi:1983sy,Sikivie:1983ip,Raffelt:1987im,Csaki:2001yk}:
\beq\label{eq:pag}
    P_0 = \frac{(2\Delta_\ag)^2}{k^2} \sin^2 \bl( \frac{k x}{2} \br) \ ,
\eeq
where $x$ is the distance traveled by the photon, and
\beqa\label{eq:oscs}
    k & \equiv & \sqrt{ (2\Delta_\ag)^2 + \bl( \Delta_a - \Delta_\gamma \br)^2 } \ , \\
    \Delta_\ag & \equiv & \frac{\gag B}{2} \ , \quad \Delta_a \equiv \frac{m_a^2}{2 \omega} \ , \quad \Delta_\gamma \equiv \frac{m_\gamma^2}{2 \omega} \ ,
\eeqa
in which $m_\gamma^2 \equiv \frac{4 \pi \alpha n_e}{m_e}$ is the effective photon mass squared in the presence of an ionized plasma with an electron number density $n_e$.\footnote{Neutral atoms, dominated by hydrogen, also contribute to the effective photon mass \cite{Born:1999ory,Mirizzi:2009iz}. For optical energies this contribution is negative but negligible, whereas for X-ray energies it is positive and sizeable. However, since the ionization is very close to 1 at the low redshifts we are interested in, this effect is subdominant when compared to the uncertainty in the value of $n_e$ itself in the IGM.}

The photons associated with typical observables travel through various environments, such as the IGM or the ICM, traversing a large number of magnetic domains. In order to quantitatively describe this phenomenon, some simplifying assumptions are made about the configuration of the magnetic fields in these environments and about the path traveled by the photons. We adopt the simple \textit{cell magnetic field} model, first introduced in \cite{Csaki:2001yk} and further developed in \cite{Grossman:2002by,Avgoustidis:2010ju}. In this model the magnetic field is assumed to be split into domains (cells) in which it can be taken to be homogeneous. The photon path, extending from a source at some distance $y$ to the observer, is assumed to cross a large number $N$ of these magnetic domains. Each {\it i}-th domain has a {\it physical} size $L_i$ and a randomly oriented magnetic field of strength $B_i$~\cite{Grossman:2002by}, whose component perpendicular to the photon's path is the same in each domain. With these simplifications, the resulting net probability of photon-axion conversion over many domains is then given by \cite{Avgoustidis:2010ju}
\beq\label{eq:pconv_prod}
    \Pag(y) = (1-A) \bl(  1 - \prod\limits_{i=1}^{N} \bl( 1 - \frac{3}{2}P_{0,i} \br) \br) \ ,
\eeq
where $A \equiv \frac{2}{3} \bl( 1 + \frac{I_a^0}{I_\gam^0} \br)$ depends on the ratio of the initial intensities of axions and photons coming from the source, denoted by $I_a^0$ and $I_\gam^0$ respectively; and $P_{0,i}$ is the conversion probability in the {\it i}-th magnetic domain, which can be obtained from \Eq{eq:pag} for $x = L_i$. 

Since $N$ is very large, \Eq{eq:pconv_prod} can be rewritten as an integral. In order to do this, we further assume that $y$ is a distance that scales linearly with $N$, such that $s = y/N$ remains constant as $N$ goes to infinity. For example, for IGM propagation the domains are typically assumed to be evenly distributed in {\it comoving} space, which means that each domain has comoving size $s$ and the distance to the source is a comoving distance $y=Ns$. Under these assumptions, we have
\beq\label{eq:pag_2}
    \Pag(y) = (1-A) \bl( 1 - \exp \bl[ \frac{1}{s} \int\limits_0^y \dd y' ~ \ln \bl( 1 - \frac{3}{2} P_0(y') \br) \br] \br) \ .
\eeq
The ratio of the observed photon flux and the emitted photon flux from the source is then given by
\beq\label{eq:psurv}
\Pgg = 1 - \Pag.
\eeq

\subsection{Intergalactic medium propagation}
\label{subsec:igm}

We will consider the propagation of photons in different media. In this section, we focus on the IGM first. The IGM, more precisely the space between large scale structures, could be home to primordial magnetic fields, which serve as ``seeds" for the observed magnetic fields in astronomical sources of different sizes, from stars to galaxy clusters. They could be generated during the preheating/reheating epochs immediately after inflation or during cosmological phase transitions before the formation of CMB. Magnetic fields produced at late times (at redshifts $z<10$) from outflows of already formed galaxies could also reside in IGM. For a review of the generation mechanisms of IGM magnetic fields, see~\cite{Durrer:2013pga}. 

At the moment, there is no direct evidence of the IGM magnetic field. Instead there are observational upper and lower bounds on the amplitude of the magnetic field in IGM. CMB anisotropies set upper limits about nG on the present value of primordial magnetic field~\cite{Trivedi_2010, Ade:2015cva, Zucca:2016iur,Paoletti:2018uic}. Other methods, such as the non-observation of Faraday rotation of the polarization plane of radio emission from distant quasars, set a similar upper limit~\cite{Durrer:2013pga}.\footnote{There is a slightly stronger upper bound on primordial magnetic field, which is $\sim 0.3-0.5$ nG from ultra-faint dwarf galaxies~\cite{Safarzadeh:2019kyq}. It is based on a strong assumption that the primordial magnetic field follows ideal magnetohydrodynamics. We will not adopt it in our paper. In~\cite{Paoletti:2018uic}, where the magnetic field is assumed to be present at the onset of recombination,  the amplitude is constrained by Planck 2015 to be smaller than $0.83\; \mathrm{nG}$.} On the other hand,  the non-observation of very high energy $\gamma$-ray cascade emission sets a lower bound on the magnetic field $B_{\rm IGM} \gtrsim 10^{-16}$G for a coherent length above Mpc and becomes more stringent at smaller coherent lengths. For recent reviews of the constraints, see~\cite{Durrer:2013pga, 2017ARA&A..55..111H, Vachaspati:2020blt}. In our paper, we will adopt the cell magnetic field model in comoving space for the IGM. More concretely, we take 1~nG as a convenient benchmark value for the component of the comoving magnetic field perpendicular to the line of sight, as well as comoving coherent length ($s_\igm$) benchmarks of $0.1~\Mpc$, $1~\Mpc$, and $10~\Mpc$ for the domain sizes. The cell magnetic field model is a very simple approximation to the structure of astrophysical magnetic fields. However, it has been shown to give the same results for axion-photon conversion as other more refined methods (such as power spectrum models) at high photon frequencies, while underestimating the conversion probability at lower frequencies \cite{Davis:2009vk,Schelpe_2010,Avgoustidis:2010ju}. The cell magnetic field model thus leads to conservative bounds on the axion parameter space. Finally, we caution the reader that the bounds we derive on axion coupling from datasets that are sensitive to the photon propagation in IGM should be understood as an upper bound on $\gag \times \frac{B_{\rm IGM}}{1\,{\rm nG}}$ for a fixed coherent length. We leave the discussion on more general combinations of $B_\igm$ and $s_\igm$ to \Sec{sec:results}.

Another important quantity of IGM that matters in our analysis is the electron density $n_e$, which determines the plasma photon mass. At low redshifts, most of the baryons are in photoionized diffuse intergalactic gas (Lyman-$\alpha$ forest) and warm-hot intergalactic matter~\cite{Nicastro:2018eam}. Among these two structures, Lyman-$\alpha$ forest contributes $28\pm11$\% of the total mass (at $z<0.5$)~\cite{Nicastro:2018eam} but occupies $\gtrsim 90\%$ of the total volume~\cite{Martizzi:2018iik}. The other structures, including warm-hot intergalactic matter, are more condensed and take up a much smaller volume. Thus, what matters more for the photon propagation is the Lyman-$\alpha$ forest. The average electron density of Lyman-$\alpha$ forest is about $6.5\times 10^{-8}$cm$^{-3}$, assuming its mass fraction to be the central value 28\%. This, however, is not the entire story. Recent simulations show that for diffuse gas, most of the volume is occupied by cosmic voids (large under-dense patches) and sheets (two-dimensional structures of matter), which constitute approximately $\sim 30\%$ and $\sim 40\%$ of the entire volume at $z\lesssim 2$ respectively \cite{Martizzi:2018iik}. Their mass fractions are significantly smaller, however, $8\%$ and $20\%$ respectively \cite{Martizzi:2018iik}.
Based on this, the electron density of the sheet component of the Lyman-$\alpha$ forest is about half of the average one over all components, $3\times 10^{-8}$cm$^{-3}$; while the electron density of the void component is about 1/4 of the average, $1.6 \times 10^{-8}$cm$^{-3}$ at $z=0$. We will take these two values as benchmarks of the plasma electron density in IGM in our analysis.

For photons traveling through the IGM,  \Eqs{eq:pag_2}{eq:psurv} could be rewritten in terms of $z$ as
\beq\label{eq:pgg_igm}
    \Pgg^{\rm IGM}(z; \bfth) = A + (1-A) \exp \bl[ \frac{1}{s} \int\limits_0^z \dd z' ~ \frac{\ln \bl( 1 - \frac{3}{2} P_0(z'; m_a, \gag) \br)}{H(z'; \Omega_\Lambda, H_0)} \br] \ , 
\eeq
where $H(z'; \Omega_\Lambda, H_0) = H_0 \sqrt{\Omega_\Lambda + (1-\Omega_\Lambda)(1+z')^3}$ is the Hubble expansion rate in flat $\LC$; and $P_0(z'; m_a, \gag)$ is the axion-photon conversion probability in \Eq{eq:pag} with all the relevant quantities appropriately rescaled by the redshift \cite{Avgoustidis:2010ju}: $s=s_\igm$ is the comoving IGM domain size; $x = L = s_\igm/(1+z')$ their physical size, $B_\igm \rightarrow B_\igm (1+z')^2$ the component of the IGM magnetic field perpendicular to the line of sight, $n_{e, \igm} \rightarrow n_{e, \igm} (1+z')^3$ the IGM electron number density, and $\omega \rightarrow \omega (1+z')$ the photon energy. The benchmark values of $B_\igm$, $s_\igm$, and $n_{e, \igm}$ are discussed and explained above.

\subsection{Intracluster medium propagation}
\label{subsec:icm}

The angular diameter distances to galaxy clusters, as we will see in \Sec{subsec:effects}, rely on measurements of cluster X-ray brightness. The X-ray photons are produced throughout the cluster via Bremsstrahlung and line-emission involving the ionized plasma composing the ICM. These photons travel first through the ICM and then the IGM to reach the detector.

Faraday rotation measurements in long wavelengths have shown \cite{deBruyn:2005ze,Taylor:2006ta,Bonafede:2010xg,Feretti:2012vk} that ICM has magnetic fields with a strength of order $\mO(\muGa)$. Therefore a fraction of the X-ray photons could convert into axions. This possibility has been studied in the literature and yields some of the strongest limits on couplings of very low mass axions to photons \cite{Wouters:2013hua, Berg:2016ese, Marsh:2017yvc,Conlon:2017qcw, Reynolds:2019uqt}. We devote the rest of this section to the computation of the effect ICM propagation has on X-rays photons as they leave the cluster. In order to perform this calculation, we need prescriptions for the ICM's electron number density $n_{e,\icm}$ and magnetic field $B_\icm$.

We model $n_{e,\icm}$ with the double-$\beta$ profile \cite{Mohr:1999ya,Bonamente:2005ct}
\beq\label{eq:neicm}
    n_{e,\icm}(r) = n_{e,0} \bl( f \bl( 1 + \frac{r^2}{r_{c1}^2} \br)^{-\frac{3\beta}{2}} + (1-f) \bl( 1 + \frac{r^2}{r_{c2}^2} \br)^{-\frac{3\beta}{2}} \br) \ ,
\eeq
where $n_{e,0}$ is the central density, $r_{c1}$, $r_{c2}$ are the two core radii, $f$ is the fractional contribution from the inner core, and $\beta$ is the slope. \Eq{eq:neicm} allows us to compute the photon plasma mass $m_\gam$, necessary for the calculation of the axion-photon conversion probability in \Eq{eq:pag}. The values of the parameters for the double-$\beta$ profiles of the clusters we use in this work can be found in \cite{Bonamente:2005ct}.

For the magnetic field we follow previous literature \cite{Bonafede:2010xg,Feretti:2012vk,Angus:2013sua,Reynolds:2019uqt} and assume the magnetic field follows a power law on the number density:
\beq\label{eq:bicm}
    B_\icm(r) = B_\reff \bl( \frac{n_e(r)}{n_e(r_\reff)} \br)^\eta \ ,
\eeq
where $r_\reff$ is some reference radius from the cluster's center, $B_\reff$ is the magnetic field value at that point, and $\eta$ some power. We will take the two models of the ICM magnetic field of the Perseus cluster found in \cite{Reynolds:2019uqt} and the one for the magnetic field of the Coma cluster in \cite{Bonafede:2010xg} as benchmarks for our analysis of the ICM effect:
\beqa
    \text{Model A}: && r_\reff = 0~\kpc, \quad B_\reff = 25~\muGa, \quad \eta = 0.7 \ , \label{eq:modA} \\
    \text{Model B}: && r_\reff = 25~\kpc, \quad B_\reff = 7.5~\muGa, \quad \eta = 0.5 \ , \label{eq:modB} \\
    \text{Model C}: && r_\reff = 0~\kpc, \quad B_\reff = 4.7~\muGa, \quad \eta = 0.5 \ . \label{eq:modC}
\eeqa
At small radii in Model A, $r < 10\;\mathrm{kpc}$, the electron number density is underestimated and spherical symmetry is unjustified. Therefore, we exclude the photon-axion conversion in the region of $r<10~\mathrm{kpc}$, following the treatment in \cite{Reynolds:2019uqt}.
We take $L_\icm=6.08~\kpc$ to be the (uniform) size of the magnetic domains, which is the mean of the $L^{-1.2}$ distribution between $3.5-10~\kpc$ proposed in \cite{Reynolds:2019uqt}.\footnote{We have also computed the line of sight-averaged survival probability by using random magnetic field domain sizes instead, drawn from the truncated power law distribution. We take one realization for each cluster, each realization in turn consisting of about $\sim 500$ domains. The corresponding constraint on $g_{a\gamma\gamma}$ is similar to our main results with at most 10\% variation. Another cross-check we have done is setting $L_\icm$ to be the lower or upper end of the truncated power law, 3.5 kpc and 10 kpc, respectively. This leads to a change in the $g_{a\gamma\gamma}$ contour within 20\% across the entire mass range. Note that we take a slightly different approach for drawing the random domains for Model B compared to \cite{Reynolds:2019uqt}. We do not account for the linear growth of the coherence length of the domain size. } We take the virial radius of the cluster to be $R_\vir = 1.8~\Mpc$, that of the Perseus cluster.\footnote{We also performed our analysis with different $R_\vir$'s for each cluster instead, using the parameters of DM halo NFW profile listed in \cite{Bonamente:2005ct}. This made the analysis more computationally expensive, and yielded identical results to those with fixed $R_\vir = 1.8~\Mpc$.} In treating the orientation of the magnetic field, we assume $B_{\rm ref}$ to be the magnetic field value in the transverse direction, perpendicular to the photon's propagation direction. If we take the direction of the magnetic field to distribute uniformly between $0$ and $\pi$, it is equivalent to substitute $B_{\rm ref}^2$ with $B_{\rm ref}^2 \left <\sin^2(\hat {\mathbf{B} } \cdot \hat{\mathbf{k}}) \right > \approx B_{\rm ref}^2/2$, with $\hat {\mathbf{B} }$ the direction of the magnetic field and $\hat{\mathbf{k}}$ that of the photon propogation. This will make the bound on $g_{a\gamma\gamma}$ derived from the ICM effect about a factor of $\sqrt{2}$ weaker. This is verified numerically by assigning a random orientation in each domain.\footnote{The code to implement the randomized magnetic field could be found  \href{https://github.com/ChenSun-Phys/cosmo\_axions}{here}.}

For X-rays originating at a radius $r$ in the cluster, we can then approximate the ratio of outgoing to initial X-ray photon flux after axion-photon conversion, following \Eq{eq:pconv_prod}, as:
\beq\label{eq:pgg_icm}
    \Pgg(r; m_a, \gag) = A + (1-A) \prod\limits_{i=1}^{N(r)} \bl( 1 - \frac{3}{2}P_0(r_i) \br) \ ,
\eeq
where $N(r) = (R_\vir-r)/L_\icm$ is the number of domains with size $L_\icm$ from origin point $r$ to the virial radius of the cluster $R_\vir$; $P_0(r_i)$ is the axion-photon probability conversion at the center $r_i$ of the {\it i}-th domain, given by \Eq{eq:pag}; $B=B_\icm(r)$ according to the three benchmarks in \Eqsto{eq:modA}{eq:modC}; and $n_e = n_{e,\icm}(r)$ according to \Eq{eq:neicm}.

Finally we want to comment on the uncertainties associated with the ICM magnetic fields. Recently, for example, just how coherent or turbulent these fields are has been the subject of some concern, and the stringent bounds on the axion-photon coupling relying on the ICM propagation in \cite{Reynolds:2019uqt} has been questioned \cite{Libanov:2019fzq}. As we will discuss below in more detail, we sidestep this issue by computing and comparing bounds based on various settings that, either include ICM photon-axion conversion with one of the benchmarks in \Eqsto{eq:modA}{eq:modC}, or ignore the ICM conversion altogether.

\subsection{Effects on distance observables}
\label{subsec:effects}
In the absence of a significant initial axion flux from the sources, non-negligible axion-photon mixing results in {\it dimming}: the brightness of a distant source will be decreased by a factor of $\Pgg$. Historically, this effect was used in an early attempt at explaining away the cosmological constant \cite{Csaki:2001yk}. While the cosmological constant has since been vindicated, the observation (or lack thereof) of dimming of distant sources can be used to constrain the axion parameter space.

In this section, we discuss the impact of dimming via axion-photon mixing on luminosity distances to type Ia SN and on angular diameter distances to galaxy clusters.

\subsubsection{Luminosity distances and distance moduli}
\label{subsubsec:ld}

The flux $F$ from a source of luminosity $L$ located at redshift $z$ is given by:
\beq\label{eq:flux}
    F(z) = \Pgg(z) \frac{L}{4\pi D_L^2(z)} \ ,
\eeq
where $\Pgg$ accounts for the possible non-conservation of photon flux between the observer and the source, and the luminosity distance $D_L$ is:
\beq\label{eq:ld}
    D_L(z) = (1+z) \int\limits_0^z \dd z' ~ \frac{1}{H(z')} \ .
\eeq

Flux measurements of distant sources such as SNIa, our primary concern in this section, are usually expressed in terms of the source's {\it apparent magnitude} $m$, which is conventionally written as:
\beqa\label{eq:mz}
    m(z) & = & M + \mu(z) \ , \nonumber \\
    \mu(z) & \equiv & -5 \log_{10} \bl( \sqrt{F(z)/F_{10}} \br) \ ,
\eeqa
where $M$ ($F_{10}$) is the absolute magnitude (flux) of the source, defined at a distance of 10 pc from it; and $\mu$ is called the {\it distance modulus} which, from \Eq{eq:flux}, can be rewritten as:
\beq\label{eq:muz}
    \mu(z) = 25 + 5 \log_{10} \bl( D_L^\eff(z)/\Mpc \br) \ ,
\eeq
where $D_L^\eff(z) = D_L(z)/\sqrt{\Pgg(z)}$ is the effective LD in the presence of axion-photon conversion, and we have taken $\Pgg$ to be 1 at a distance of 10 pc from the source.

Putting everything together and making explicit the dependence on the parameters $\bfth$ in our analysis, the effective apparent magnitude of the SNIa located at redshift $z$ is
\beqa\label{eq:meff}
    m^\eff(z; \bfth, M) & = & M + 25 + 5 \log_{10} \bl( D_L^\eff(z; \bfth)/\Mpc \br) \ , \\
    D_L^\eff(z; \bfth) & = & D_L(z; \Omega_\Lambda, H_0)/\sqrt{\Pgg(z; \bfth)} \ ,
\eeqa
with $D_L(z; \Omega_\Lambda, H_0)$ given by \Eq{eq:ld} and $\Pgg(z; \bfth)$ by \Eq{eq:pgg_igm}. We will take $A$ in \Eq{eq:pgg_igm} to be 2/3 since the initial axion flux from SNIa is negligible~\cite{Grossman:2002by}. Note that Ref.~\cite{Grossman:2002by} didn't consider the possibility of resonant production of axions at SNIa, which we will entertain in App.~\ref{app:A} and show that it doesn't modify the conclusion.

\begin{figure}[th]
  \centering
  \includegraphics[width=.6\textwidth]{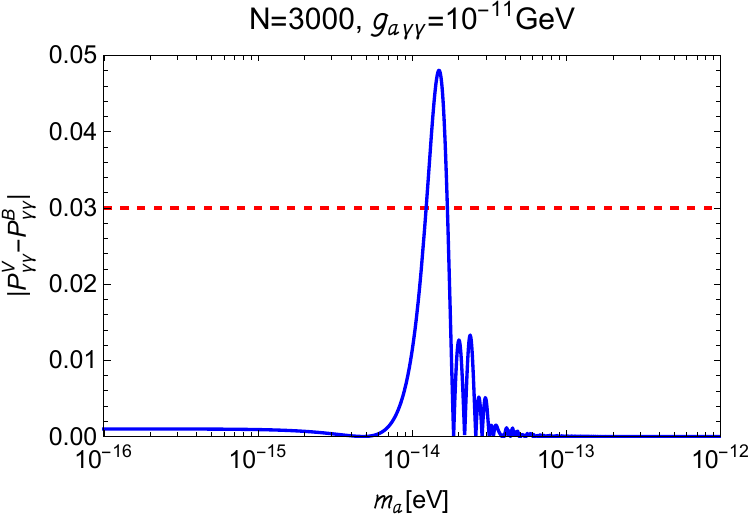}
  \caption{The achromaticity bound assuming 3000 random magnetic field domains and $g_{a\gamma\gamma}  = 10^{-11} \;\mathrm{GeV}$. }
  \label{fig:color-constraint}
\end{figure}

Finally, we want to comment on the energy dependence of the photons. The photons from the SNIa are in the optical band with $\omega \approx 1~\eV$. If the source is observed in various frequencies, its magnitude or flux will in general undergo spectral distortion (also called chromaticity) as a result of the photon energy dependence of $\Pgg$, which can in principle be used to further constrain the axion parameter space. In the parameter space we are interested in, however, this distortion is negligible.
This can be estimated by comparing the oscillation probabilities for the B band ($4.3~\mathrm{eV}$) and V band ($3.4~\mathrm{eV}$) photons respectively. The achromaticity requirement from data~\cite{Perlmutter:1998np} could be translated into a constraint of $\left|P_{\gamma\gamma}^B - P_{\gamma\gamma}^V\right| \lesssim 0.03$ \cite{Csaki:2001jk}. For illustrative purposes let us assume that the photons transverse $N=3000$ magnetic domains and $\gag = 10^{-11}$ GeV$^{-1}$. The probability difference computed using \Eq{eq:pag_2} and \Eq{eq:psurv} is presented in Fig.~\ref{fig:color-constraint}. One can see from the figure that even with this relatively large $g_{a\gamma\gamma}$ (which is already excluded by SN1987a~\cite{Payez:2014xsa}), the monochromaticity requirement only constrains $m_a$ in a tiny range around $10^{-14}$ eV. Therefore, we do not consider achromaticity in the SNIa observations from photon-axion conversion further in our analysis.

\subsubsection{Angular diameter distances of galaxy clusters}
\label{subsubsec:modifiedadd}

The angular diameter distance $D_A$ (ADD) is defined as the ratio of an astrophysical object's physical size $d$ to the arc $\theta$ that it subtends in the sky. It can be shown to be equal to:
\beq\label{eq:add}
    D_A(z) = \frac{d}{\theta} = \frac{1}{1+z} \int\limits_0^z \dd z' ~ \frac{1}{H(z')} \ .
\eeq

In general ADDs are unaffected by the axion-photon conversion since they do not rely on brightness measurements of any kind. This is the case, for example, for observations that measure the imprint of the comoving sound horizon on the galaxy two-point correlation function, such as the BAO measurements. However, microwave and X-ray surveys can be used to determine ADDs to galaxy clusters, which would then be impacted by axion-photon conversion. Indeed, it has been shown in~\cite{Bonamente:2004vx} that ADDs to galaxy clusters can be obtained from measurements of the clusters' X-ray surface brightness $S_\mathrm{X}$, due to  Bremsstrahlung and line-emission resulting from ion-electron collisions in the ICM; combined with observations of the brightness temperature decrement $\Delta T_\mathrm{SZ}$ from CMB photons undergoing inverse Compton scattering with the same ICM, the so-called Sunyaev-Zeldovich effect (SZE)\footnote{Note that any prior photon-axion conversion effects in IGM, as the photons travel from the surface of last scattering to the clusters, would leave the brightness temperature decrement $\Delta T_\mathrm{SZ}$ unaffected. This is because this quantity is concerned with the relative brightness of the CMB photons that pass through clusters compared to those that do not. However, the \textit{absolute} value of CMB distortion effects due to photon-axion conversion from the surface of last scattering can be used to constrain the axion parameter space, see \cite{Mirizzi:2005ng,Mirizzi:2009nq}.}:
\beq\label{eq:add_cls}
    D_A(z) \propto \frac{\Delta T_\mathrm{SZ}^2}{S_\mathrm{X} (1+z)^4} \ .
\eeq

As we have seen in \Sec{subsec:icm}, X-rays originate in the cluster's ICM and are affected by axion-photon conversion as they travel through the ICM magnetic fields. \Eq{eq:pgg_icm} describes the ratio of outgoing to produced X-rays flux, for photons traveling radially outwards from their oirigin at a distance $r$ from the cluster's center. Since clusters are extended objects in the sky, the change in the X-ray brightness $S_\mathrm{X}$ is not exactly described by \Eq{eq:pgg_icm}, there being photons whose path from the interior to the exterior of the cluster and from there to the observer is not radial. Nevertheless, we expect that a good proxy for the exact ICM effect across the surface of the cluster is to weigh $\Pgg$ by the integrand sourcing the brightness $S_\mathrm{X}$ and average over the line of sight. Indeed, $S_\mathrm{X} \propto \int \dd l ~ n_{e,\icm}^2 \Lambda_{ee}$ \cite{Bonamente:2004vx}, where $l$ is the line-of-sight variable, and $\Lambda_{ee}$ the cluster cooling rate, which scales like the square root of the ICM temperature, $\sim T_e^{1/2}$. From \cite{Bonamente:2005ct}, we see the ICM temperature is approximately constant throughout most clusters, except for only a handful of them where the temperature fluctuates at most by a factor of four with small angular variation, such as RX J1347.5-1145, Abell 1835, Abell 2204, and Abell 1914. The resulting factor of two change in $\Lambda_{ee}$ leads to a percentage level change in the weighted average of $P_{\gamma\gamma}$. Therefore, we approximate the suppression on the X-ray brightness $S_\mathrm{X}$ due to the ICM effect with:
\beq\label{eq:avgPggicm}
    \avgPggicmpars \equiv \frac{\int\limits_{r_\ini}^{R_\vir} \dd r ~ n_{e,\icm}^2(r) \Pgg(r; m_a, \gag)}{\int\limits_{r_\ini}^{R_\vir} \dd r ~ n_{e,\icm}^2(r)} \ ,
\eeq
where the integral is taken from some initial radius $r_\ini$. For the ICM magnetic field Model A we follow \cite{Reynolds:2019uqt} and take $r_\ini = 10~\kpc$, whereas for models B and C we take $r_\ini=0~\kpc$.

The suppression described in \Eq{eq:avgPggicm} immediately implies that there is a fraction $1-\avgPggicm$ of the initial X-ray flux that has converted into axions. This means that the ratio of axions to photons outside the cluster is given by:
\beq\label{eq:icm_intens}
    \frac{I_a^\mathrm{clusters}}{I_\gam^\mathrm{clusters}} = \frac{1-\avgPggicm}{\avgPggicm} \ .
\eeq
This changes the value of $A_\mathrm{X} = \frac{2}{3}\bl( 1 + \frac{I_a^\mathrm{clusters}}{I_\gam^\mathrm{clusters}} \br)$ in \Eq{eq:pgg_igm} that describes the subsequent X-ray propagation in the IGM.

Following the scaling described in \Eq{eq:add_cls}, we can combine \Eqs{eq:avgPggicm}{eq:icm_intens} with the formula in \Eq{eq:pgg_igm}, for photons of both CMB and X-ray energies propagating in the IGM, in order to finally arrive at the effective ADDs to clusters:
\beq\label{eq:DAeff}
    D_A^\eff(z; \bfth) = D_A(z; \Omega_\Lambda, H_0) ~ \frac{\Pgg^{\rm IGM}(z; \bfth, \omega_\mathrm{CMB})^2}{\Pgg^{\rm IGM}(z; \bfth, \omega_\mathrm{X}, A_\mathrm{X})\avgPggicmpars} \ ,
\eeq
with $D_A(z; \Omega_\Lambda, H_0)$ given by the standard cosmology formula in \Eq{eq:add}. $A_\mathrm{CMB} = \frac{2}{3}$ is used in the computation of the numerator in \Eq{eq:DAeff} according to \Eq{eq:pgg_igm}, since the ICM only affects photons of microwave energy in a negligible way.

\section{Data and methodology}
\label{sec:data}

Having explained described the effects that axion-photon conversion has on various cosmological observables, we devote this section to describing the datasets and methodology we have used for our model fits.

\subsection{Datasets}
\label{subsec:data}

For our analysis we consider data from the following experiments, which we will combine in different ways:
\begin{itemize}
    \item \textbf{Pantheon}: the Pantheon dataset \cite{Scolnic:2017caz}, consisting of apparent magnitude measurements of 1048 SNIa;
    \item \textbf{Clusters}: a set of ADDs measurements for 38 galaxy clusters \cite{Bonamente:2005ct};
    \item \textbf{SH0ES}: measurements of the absolute magnitudes of 19 SNIa by the SH0ES collaboration \cite{Riess:2016jrr};
    \item \textbf{TDCOSMO}: the Hubble parameter as measured by the TDCOSMO collaboration using strong lensing \cite{Birrer:2020tax};
    \item \textbf{BAO}: the measurements of the imprint of baryon acoustic oscillations in galaxy distributions \cite{Beutler:2011hx,Ross:2014qpa,Alam:2016hwk}; 
    \item \textbf{Planck}: The value of the comoving sound horizon at the epoch of baryon drag, given by the Planck collaboration's observation of CMB anisotropies \cite{Aghanim:2018eyx}.
\end{itemize}

In the rest of this section, we will describe in more detail these datasets and provide their corresponding likelihoods, some of which are inspired by {\tt MontePython}~\cite{Audren:2012wb,Brinckmann:2018cvx}. We will then use these likelihoods to constrain the axion parameter space. 

\subsubsection{Pantheon}
\label{subsubsec:pantheon}

As we have seen in the previous section, axion-photon conversion impacts those cosmic distance measurements that rely on the brightness of astrophysical sources. The observation of the brightness of SNIa is one such kind of measurement. The Pantheon dataset is the most up-to-date collection of apparent magnitude measurements for 1048 SNIa in the redshift range of $0.01 < z < 2.3$ \cite{Scolnic:2017caz}. The corresponding likelihood we use is given by
\beqa
    -2 \ln \mL_\Pan & = & \sum\limits_{i,j=1}^{1048} \Delta_i C_{ij}^\Pan \Delta_j \ , \\
    \Delta_i & \equiv & m_i^\Pan - m^\eff (z_i; \bfth, M) \ ,
\eeqa
where $C^\Pan$ is the Pantheon inverse covariance matrix; $m_i^\Pan$ is the Pantheon measurements for the apparent magnitudes of the SNIa located at redshift $z_i$ while $m^\eff(z_i; \bfth, M)$ is the corresponding theory prediction given by \Eq{eq:meff} in the axion-photon conversion model. We take $M$ as a free parameter in our MCMC runs and fit together with the model parameters.

For the SNIa in the Pantheon set, we will take the energy of their optical photons to be $\omega = 1~\eV$, the IGM magnetic field $B_\igm = 1~\nGa$, the comoving size of the magnetic fields $s_\igm \in \{ 0.1, 1, 10\}~\Mpc$, and the IGM electron number density $n_{e,\igm}$ either $1.6 \times 10^{-8}~\cm^{-3}$ or $3.0 \times 10^{-8}~\cm^{-3}$. All benchmarks are in accordance with the discussion in \Sec{subsec:igm}.

\subsubsection{Cluster angular diameter distances}
\label{subsubsec:add}

Measurements of angular diameter distances (ADDs) to galaxy clusters, inferred from SZE and X-ray cluster data, are also sensitive to axion-photon conversion in a manner described in \Sec{subsec:effects}, and can therefore be used to constrain the axion parameter space. In our present work we use the sample of 38 clusters from \cite{Bonamente:2005ct} as listed in their Table 2, which assumes spherically symmetric clusters in hydrostatic equilibrium.\footnote{The sphericity requirement is relaxed in the sample of 25 clusters studied in \cite{DeFilippis:2005hx}, where an elliptical morphology is assumed instead. In it, however, the values of the $\LC$ cosmological parameters are fixed, since they are highly degenerate with the shape parameters, whose determination is the main goal of the paper. Since we are interested in fitting the cosmological parameters along with those of the axion-photon system, we use the dataset in \cite{Bonamente:2005ct} instead. Note that \cite{Bonamente:2005ct} quantifies an error of 15\% arising from the sphericity assumption.}

We then construct a likelihood for the ADD measurements from this dataset taking into account the statistical and systematic uncertainties enumerated in Table 3 of \cite{Bonamente:2005ct}, which we add in quadrature. The likelihood is given by:
\beq\label{eq:add_lkl}
    -2\ln \mL_\cl = \sum\limits_{i=1}^{38} \bl( \frac{D_{A,i}^\cl - D_A^\eff(z_i; \bfth)}{\sigma_i^\cl} \br)^2 \ ,
\eeq
where $D_A^\eff(z_i; \bfth)$ is given by \Eq{eq:DAeff}. The data in \cite{Bonamente:2005ct} provides not only the redshifts and ADDs to these clusters but also the means and uncertainties of the $n_{e,0}$, $f$, $r_{c1,c2}$, and $\beta$ parameters for the double-$\beta$ profile of \Eq{eq:neicm} describing the ICM electron number density $n_{e,\icm}$. We have computed the impact of the uncertainties in these parameters on the line of sight-averaged ICM survival probability of \Eq{eq:avgPggicm}. We found that these uncertainties only lead to a variation below 4\% on top of the result using the mean values. We therefore ignore these subdominant effects, and restrict ourselves to the mean values provided by \cite{Bonamente:2005ct}.

For the factors in \Eq{eq:DAeff} related to IGM propagation we assume the same benchmark quantities as for the SNIa Pantheon dataset. For the factors dealing with the ICM effect, we use the three benchmark magnetic field models described in \Eqsto{eq:modA}{eq:modC}. We take the CMB photons to have energy $\omega_\mathrm{CMB} = 2.4 \times 10^{-4}~\eV$. We average the X-ray photon energy in the band $0.7 - 7\; \mathrm{keV}$ using the measured temperature of each cluster \cite{Bonamente:2005ct}, and the resulting photon effective energy is around $\omega_\mathrm{X} = 5\; \mathrm{keV}$, which we use for our fits. In light of the uncertainties in the axion-photon conversion for X-rays in the ICM discussed in \Sec{subsec:icm}, we also perform fits to the ADD data ignoring the ICM effect.

\subsubsection{BAO}
\label{subsubsec:bao}

Galaxy surveys can determine the imprint of baryon acoustic oscillations on matter distribution and then ADDs at low redshifts. More concretely, they measure ratios of the comoving sound horizon at the epoch of baryon drag $\rsd$ to either the comoving angular diameter distance $D_M(z) \equiv (1+z)D_A(z)$, the Hubble distance $D_H(z) \equiv z/H(z)$, or the combined distance $D_V(z) \equiv \bl( D_M(z)^2 D_H(z) \br)^{1/3}$.

We use the recent observations of $\rsd/D_V$ at $z = 0.106$ by 6dFGS \cite{Beutler:2011hx}, of $D_V/\rsd$ at $z = 0.15$ by SDSS using the MGS galaxy sample \cite{Ross:2014qpa}, and of both $D_M/\rsd$ and $\rsd/D_H$ at $z=0.38, ~ 0.51, ~\text{and}~ 0.61$ by BOSS, from the CMASS and LOWZ galaxy samples of SDSS-III DR12 \cite{Alam:2016hwk}. We use the covariance matrix to take care of the correlation between the three redshift bins from BOSS as the middle one completely overlaps with the other two. There is no correlation between 6dFGS, MGS sample, and BOSS since BOSS only contains data with $z> 0.2$.

Note that since none of these surveys rely on the brightness of sources, these measurements are insensitive to axion-photon conversion effects, and therefore can be used to constrain the cosmological parameters $\{ H_0, \Omega_\Lambda \}$ of $\LC$. Since these measurements depend on $\rsd$, whenever we use these datasets we include $\rsd$ as an extra parameter to our model.

Schematically, then, the BAO likelihood is given by:
\beqa\label{eq:bao}
    -2 \ln \mL_\BAO & = & \sum\limits_{i,j} \Delta_i C_{ij}^\BAO \Delta_j \ , \\
    \Delta_i & \equiv & Q^\BAO_i - Q^{\LC}(z_i; \Omega_\Lambda, H_0, \rsd) \ ,
\eeqa
where $C^\BAO$ is the inverse covariance matrix of the BAO measurements. $Q^\BAO_i$ is the quantity being measured at redshift $z_i$, and $Q^{\LC}(z_i; \Omega_\Lambda, H_0, \rsd)$ is the model's prediction, which depends only on the cosmological parameters $\{ H_0, \Omega_\Lambda, \rsd \}$ and is therefore identical to that of $\LC$.

\subsubsection{SH0ES}
\label{subsubsec:shoes}

The SH0ES collaboration used parallax to deduce the distances to standard candles such as Cepheid variables in order to determine the absolute magnitude of 19 accompanying SNIa \cite{Riess:2016jrr}. We then construct the corresponding likelihood:
\beq\label{eq:shoes}
    -2 \ln \mL_\SHOES = \sum\limits_{i=1}^{19} \bl( \frac{M_i^\SHOES - M}{\sigma_i^\SHOES} \br)^2 \, .
\eeq
Note that we are using the SH0ES collaboration's determination of the absolute magnitude $M$ and not their value for $H_0$, since this was determined under the assumption of photon flux conservation, which is not true in the axion-photon conversion framework.

\subsubsection{TDCOSMO}
\label{subsubsec:h0}

The Hubble parameter can be determined through strong lensing. A sample of 7 such lenses was used by the TDCOSMO collaboration to determine a value of $H_0 = 74.5^{+5.6}_{-6.1} ~ \km \; \sec^{-1} \Mpc^{-1}$ \cite{Birrer:2020tax}. We note that this measurement is independent of photon brightness and thus constrains $H_0$ only, not the axion parameter space. The likelihood we use is therefore:
\beq
    -2 \ln \mL_\TDCOSMO = \bl( \frac{H_0^\TDCOSMO - H_0}{\sigma^\TDCOSMO} \br)^2 \ ,
\eeq
where for simplicity we take the symmetrized error $\sigma^\TDCOSMO = 5.85~\km \; \sec^{-1} \Mpc^{-1}$.

\subsubsection{Planck}
\label{subsubsec:early}

The use of the BAO likelihood defined in \Eq{eq:bao} requires the introduction of the comoving sound horizon at baryon drag $\rsd$ as an extra parameter in our model. There is enough constraining power in the late-Universe data from Pantheon+SH0ES+TDCOSMO to determine the value of this parameter. However a different possibility is to use early-Universe data from Planck's observations of the CMB anisotropies \cite{Aghanim:2018eyx}, which yield $r_s^\mathrm{drag,obs} = 147.09 \pm 0.26$ for TT,TE,EE+low-E+lensing measurements. The likelihood we use is then simply given by:
\beq
    - 2 \ln \mL_\Planck = \bl( \frac{r_s^\mathrm{drag,Pl} - \rsd}{\sigma^\Planck} \br)^2 \ .
\eeq
In the next section we describe how we deal with the so-called Hubble crisis and the discrepancies between SH0ES, TDCOSMO, and Planck.

\subsection{Methodology}
\label{subsec:method}

We consider various combinations of the datasets described in \Sec{subsec:data}, as well as different assumptions regarding the ICM and IGM, with the goal of deriving and comparing bounds on the axion parameter space $( m_a, \gag )$ in different cases:
\begin{itemize}
    \item {\it Early vs. Late}: In light of the Hubble crisis and the disagreement regarding $H_0$ and $\rsd$ between the SH0ES collaboration on the one hand and the Planck collaboration on the other \cite{Aylor:2018drw,Knox:2019rjx}, we split our datasets into two subsets with likelihoods given by
    \beqa\label{eq:early}
        \mL_\early & \equiv & \mL_\Pan \cdot \mL_\cl \cdot \mL_\BAO \cdot \mL_\Planck \ ,\\
        \mL_\late & \equiv & \mL_\Pan \cdot \mL_\cl \cdot \mL_\BAO \cdot \mL_\SHOES \cdot \mL_\TDCOSMO  \ ,\label{eq:late}
    \eeqa
    and we fit to each likelihood separately.
    
    \item {\it with vs. without ICM propagation}: Given the debate surrounding the robustness of bounds on axion-photon interactions obtained from ICM propagation effects \cite{Reynolds:2019uqt,Libanov:2019fzq}, we perform fits both with and without this effect. For the analyses that include the ICM effect, we assume three different models for the ICM magnetic field: A, B, and C, given by \Eqsto{eq:modA}{eq:modC}.
    
    \item {\it IGM electron number density}: We take two benchmarks for the IGM electron number density: $n_{e,1} = 1.6 \times 10^{-8} ~\cm^{-3}$ and $n_{e,2} = 3.0 \times 10^{-8}~\cm^{-3}$, described in \Sec{subsec:igm}.
    
\item {\it IGM magnetic domain sizes}: We take three benchmarks for the coherent length of the IGM magnetic domains: $s_\igm \in \{ 0.1, 1, 10 \}~\Mpc$, described in \Sec{subsec:igm}. 
\end{itemize}

We employ the likelihood-ratio test in order to find the 95\% confidence level (C.L.) one-sided upper limits on the axion parameter space, putting constraints on the axion's coupling to photons at fixed masses. We will describe the procedure below, following Ref.~\cite{Cowan:2010js}. We first take the total likelihood $\mL_\tot(\bfTH)$ from one of the options described in the itemized cases above, based on either \Eq{eq:early} or \Eq{eq:late}, for each of the ICM and IGM assumptions listed above. 
$\mL_\tot(\bfTH)$ is a function of the combined set of cosmological and axion parameters $\bfTH = \{ H_0, \Omega_\Lambda, M, \rsd, m_a, \gag \}$. We then scan over the parameter space of $\bfxi \equiv \{H_0, \Omega_\Lambda, M, r_s^\mathrm{drag}, \gag\}$ to find the maximal likelihood \textit{at a fixed axion mass} $m_a$, $\maxim_{\bfxi}\; \mL_\tot(m_a) \equiv \mathcal{L}_\maxim (m_a)$. We allow $\gag$ to take imaginary values as well, in order to cover the case with negative signal strength $\gag^2 < 0$. We then follow the prescription in Ref.~\cite{Cowan:2010js} to construct the test statistic in \Eq{eq:test_stat}. Instead of taking an evenly spaced grid, we make use of the public MCMC code {\tt emcee}~\cite{Foreman_Mackey_2013} to speed up the maximization of the likelihood.
After that, at each point in the $\{m_a, g_{a\gamma\gamma}\}$ axion parameter space, we scan over the set of cosmological parameters only $\bfa \equiv \{ H_0, \Omega_\Lambda, M, \rsd \}$ to find $\maxim_{\bfa} \; \mL_\tot \equiv \mathcal{L}_\maxim (m_a, \gag)$. Then we compute the log-likelihood ratio, 
\beq\label{eq:test_stat}
\Delta \chi^2_\tot \equiv -2 \ln  [\mathcal{L}_\maxim(m_a, \gag)/\mathcal{L}_\maxim(m_a)] \, .
\eeq
We then set $\Delta \chi^2_\tot = 2.71$ (the value for 95\% C.L. one-sided upper limits with one degree of freedom \cite{Cowan:2010js}) to find the 95\% C.L. upper bound. The range of parameters we scan is: $0.6 < \Omega_\Lambda < 0.75,\; 60< H_0/(\km \; \sec^{-1} \Mpc^{-1}) <  80,\; -21 < M <  -18,\; 120~\Mpc < \rsd <  160~\Mpc,\; 10^{-17} < (m_a/\eV) < 10^{-11},\; 10^{-18} <(\gag \GeV) < 10^{-8}$. This range is motivated by physical considerations (e.g., $\gag^{-1} < 10^{18}$ GeV, the Planck scale)\footnote{Note that when $\gag$ is sufficiently small, i.e., close to $10^{-18}$ GeV$^{-1}$, the axion-photon conversion is negligible and the model is essentially $\Lambda$CDM.} and sufficiently broad to include the points that maximize the likelihoods.

\section{Results}
\label{sec:results}

In this section, we will present our results, based on the datasets and methodology described in the previous section. Our {\tt python} code, which implements both the physics of axion-photon conversion and the MCMC analysis of its parameters using {\tt emcee}, is publicly available at \href{https://github.com/ManuelBuenAbad/cosmo_axions}{{\tt github.com/ManuelBuenAbad/cosmo\_axions}}. 

First, we vary both $m_a$ and $\gag$ to find the best fits for a given dataset. We find no evidence for axion-photon conversions, for any combinations of datasets, with the IGM and ICM assumptions considered and listed in the previous section. In order to investigate whether the axion-photon conversion and its impact on physical observables has been mismodelled, or whether there are unaccounted systematics in our datasets, we need to compare the best fit $\chi^2$ values with non-negative signal strengths (\ie real $\gag$) to those where negative signal strengths (\ie imaginary $\gag$) are allowed. When only non-negative signal strengths are considered, we find that the best fits are consistent with the null hypothesis, or equivalently, $\Lambda$CDM with no axion-photon conversion. Now we allow negative signal strengths.
For our analyses involving ${\cal L}_{\rm early}$ and no ICM conversion, $\gag$'s at best fits take imaginary values with the absolute values in the range between $10^{-12}\text{--}10^{-11}~\GeV^{-1}$, and a $|\Delta \chi^2|$ difference from that of the null hypothesis in the range $3.2\text{--}3.8$. For cases with ${\cal L}_{\rm early}$ and ICM conversion, $\gag$'s at best fits are around imaginary $10^{-12}~\GeV^{-1}$, with $|\Delta \chi^2| \lesssim 2.2$ with respect to the null hypothesis. On the other hand, fitting with ${\cal L}_{\rm late}$ yields $|\gag| \sim 10^{-13}\text{--}10^{-11.5}~\GeV^{-1}$ (some real, some imaginary) with $|\Delta \chi^2| < 1$, independent of whether ICM conversion is implemented or not. In most of these cases, the $|\Delta \chi^2|$ difference between the best fits allowing only non-negative signal strengths and those allowing negative signal strengths is below $1\sigma$. In all of the cases, this difference lies below the $2\sigma$ level (the two-sided $\chi^2$ thresholds for two degrees of freedom are $2.28$ and $5.99$ for 68\% and 95\% respectively). This indicates both a consistent modeling of the axion-photon conversion and a correct accounting of the uncertainties in the dataset. 

\begin{figure}[h]
  \centering
  \includegraphics[width=0.495\textwidth]{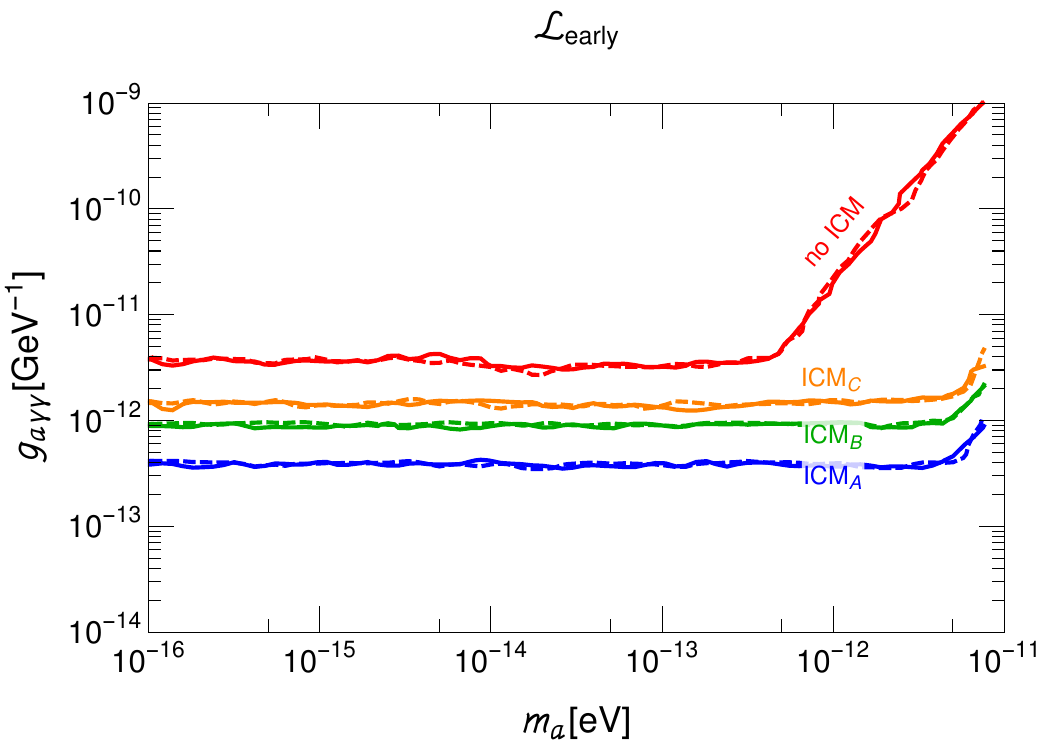}   \includegraphics[width=0.495\textwidth]{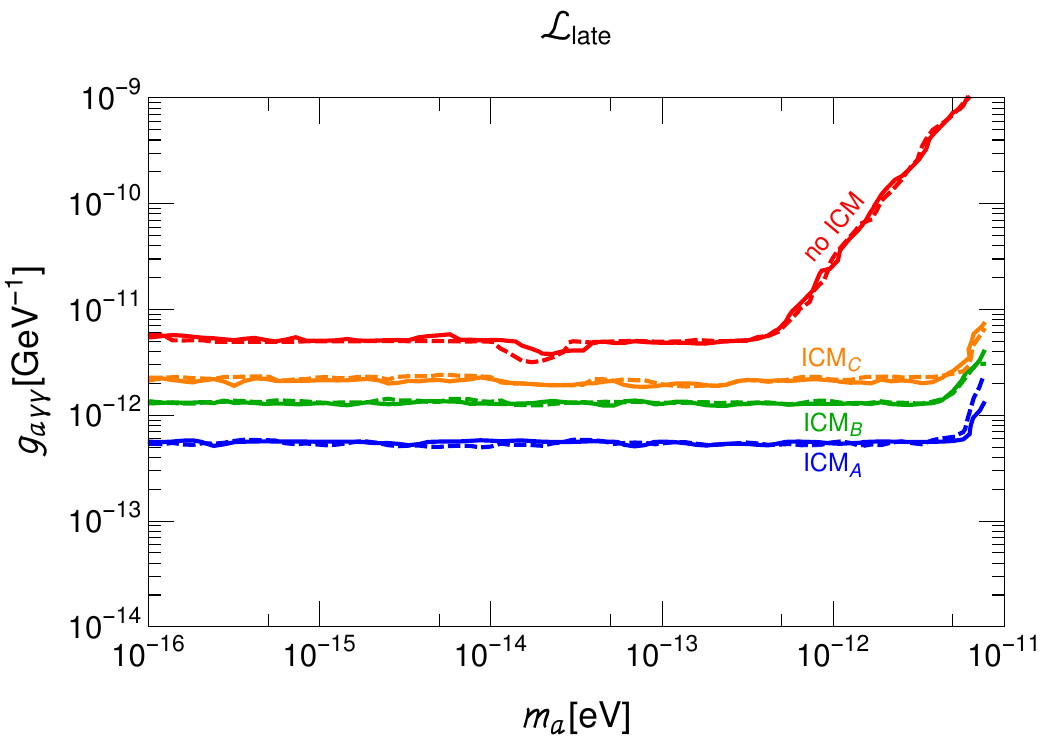}
  \caption{95\% C.L. upper bound on $\gag$ as a function of $m_a$ from likelihood-ratio tests. We assume $B_{\rm IGM}=1$ nG. Left: bound from ${\cal L}_{\rm early}$; Right: bound from ${\cal L}_{\rm late}$. Dashed curves assume $n_{e,1} = 1.6 \times 10^{-8}$ cm$^{-3}$ while solid curves assume $n_{e,2} = 3.0 \times 10^{-8}$ cm$^{-3}$. From top to bottom, the four sets of curves (each set with a solid and a dashed line for two different $n_e$'s) correspond to not including ICM effects on the galaxy cluster data, or including ICM effects assuming magnetic field models A, B, or C in \Eqsto{eq:modA}{eq:modC}. Here we set $s_\igm = 1~\Mpc$. The tiny differences between the solid and dashed lines are numerical noise, as we have corroborated with multiple runs.
  \label{fig:chisq}
}
\end{figure}

We then use the log-likelihood ratio in Eq.~\eqref{eq:test_stat} to obtain 95\% C.L. upper bound in the $(m_a, \gag)$ plane assuming $B_{\rm IGM} = 1$ nG from both ${\cal L}_{\rm early}$ and ${\cal L}_{\rm late}$, which is shown in \Fig{fig:chisq}. We also show constraints from varying the electron density in IGM and models to describe possible ICM effects on $D_A$ to galaxy clusters as described in \Sec{subsec:icm} and \Sec{subsubsec:modifiedadd}.

\begin{figure}[h]
  \centering
  \includegraphics[width=0.8\textwidth]{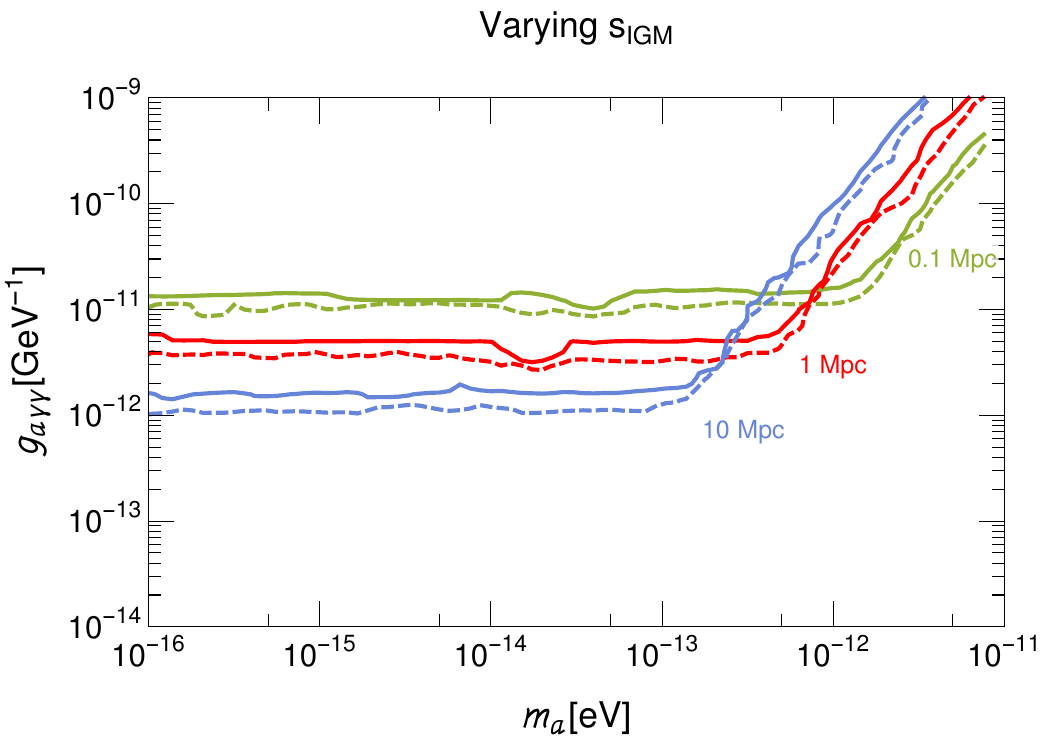}
  \caption{95\% C.L. upper bounds on $\gag$ as a function of $m_a$ from likelihood-ratio tests, for various coherent lengths of IGM magnetic field. Dashed curves are from ${\cal L}_{\rm early}$, while solid curves are from ${\cal L}_{\rm late}$. We have taken $B_{\rm IGM}=1$ nG and $n_{e,1} = 1.6 \times 10^{-8}$ cm$^{-3}$. We do not include ICM effects on the galaxy cluster data here.
  \label{fig:varysIGM}
}
\end{figure}

From all the numerical results, we learn that
\begin{itemize}
    \item The results are very similar for both ${\cal L}_{\rm early}$ and ${\cal L}_{\rm late}$, for a given ICM model. The datasets used in ${\cal L}_{\rm early}$ and ${\cal L}_{\rm late}$ mainly differ in $H_0$ and $r_s^{\rm drag}$ anchors for BAO. Yet the constraints on $\gag$ are mainly due to the shape of $H(z)$ at late times constructed from various distance measurements, which are used for both ${\cal L}_{\rm early}$ and ${\cal L}_{\rm late}$. In other words, {\it our bounds do not depend on the resolution of Hubble crisis.}
    
    \item The bounds are insensitive to the precise values of $n_{e,\igm}$.
    
    \item We assume $B_\igm = 1$ nG. For our results {\it without} ICM effects (red curves in all the figures of this section), the bounds should be understood as being constraints on $\gag \times \frac{B_\igm}{1 \,\nGa}$, for fixed values of the IGM coherent length $s_\igm$. A better understanding of $B_\igm$ could help improve these bounds.
    
  \item Increasing the coherent length $s_\igm$ strengthens (weakens) the constraints for small (large) axion masses in proportion to $\propto 1/\sqrt{s_\igm / 1 ~ \Mpc}$ ($\propto \sqrt{s_\igm / 1 ~ \Mpc}$). The reason for this can be understood from a heuristic treatment of \Eqs{eq:pag}{eq:pag_2}. For small axion masses, the conversion probability $P_0$, within a single IGM magnetic domain, is in its linear regime, \ie $k \ll s_\igm$ in \Eq{eq:pag} and $P_0 \sim (\gag ~ B_\igm ~ s_\igm)^2$. \Eq{eq:pag_2} in turn implies $\Pag \sim y ~ s_\igm ~ (\gag ~ B_\igm)^2$ for sources at a comoving distance $y$. Consequently the bounds on $\gag$ scale with $\propto 1/\sqrt{s_\igm}$ for small axion masses. On the other hand, for large axion masses $P_0$ oscillates rapidly within a single IGM domain and $\sin^2(kx/2)$ averages to $1/2$. Therefore $P_0 \sim (\gag ~ B_\igm ~ \omega)^2 / m_a^4$ and $\Pag \sim \frac{y}{s_\igm} (\gag ~ B_\igm ~ \omega)^2 / m_a^4$. This means that the bounds on $\gag$ scale with $\propto \sqrt{s_\igm}$ for large axion masses. The transition between both regimes occurs when $k x \sim \mO(1)$, \ie at $m_a \approx \sqrt{\omega / s_\igm}$.
    
    \item Including the ICM photon-axion conversion effects on the X-ray propagation used for inferring $D_A$ to galaxy clusters will make the upper limit stronger, for all the ICM magnetic field models we consider. In particular, for Model A in \Eq{eq:modA}, the upper bound on $\gag$ could be improved by one order of magnitude compared to the bound assuming no ICM effect. What is more, these ICM effects completely overshadow those from IGM propagation. Even choosing as a benchmark the smallest possible IGM magnetic field, $B_\igm = 10^{-16}~\Ga$, the constraints from ICM conversions remain the same, as strong as the ones presented in \Fig{fig:chisq}. In other words, \textit{constraints that include ICM effects are independent of the magnetic field strength of IGM}.
    
    \item The galaxy cluster ADD measurements drive the likelihood-ratio, with subdominant contributions from the Pantheon dataset, both in the case where we ignore or include ICM X-ray conversion effects. The Pantheon SNIa dataset by itself does place constraints in the ($m_a$, $\gag$) parameter space, albeit somewhat weaker ones.
\end{itemize}

\begin{figure}[h]
  \centering
  \includegraphics[width=0.495\textwidth]{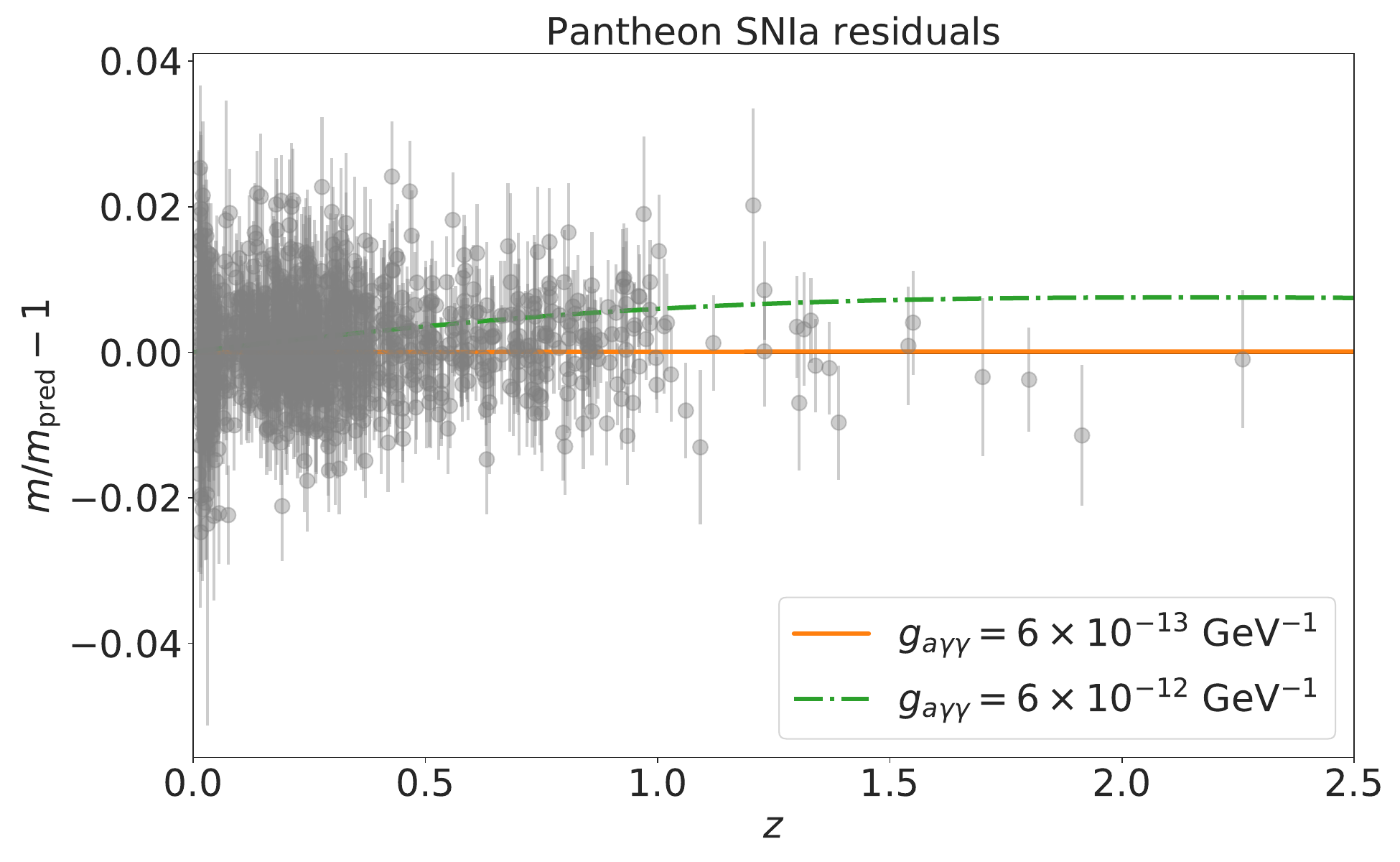}
  \includegraphics[width=0.495\textwidth]{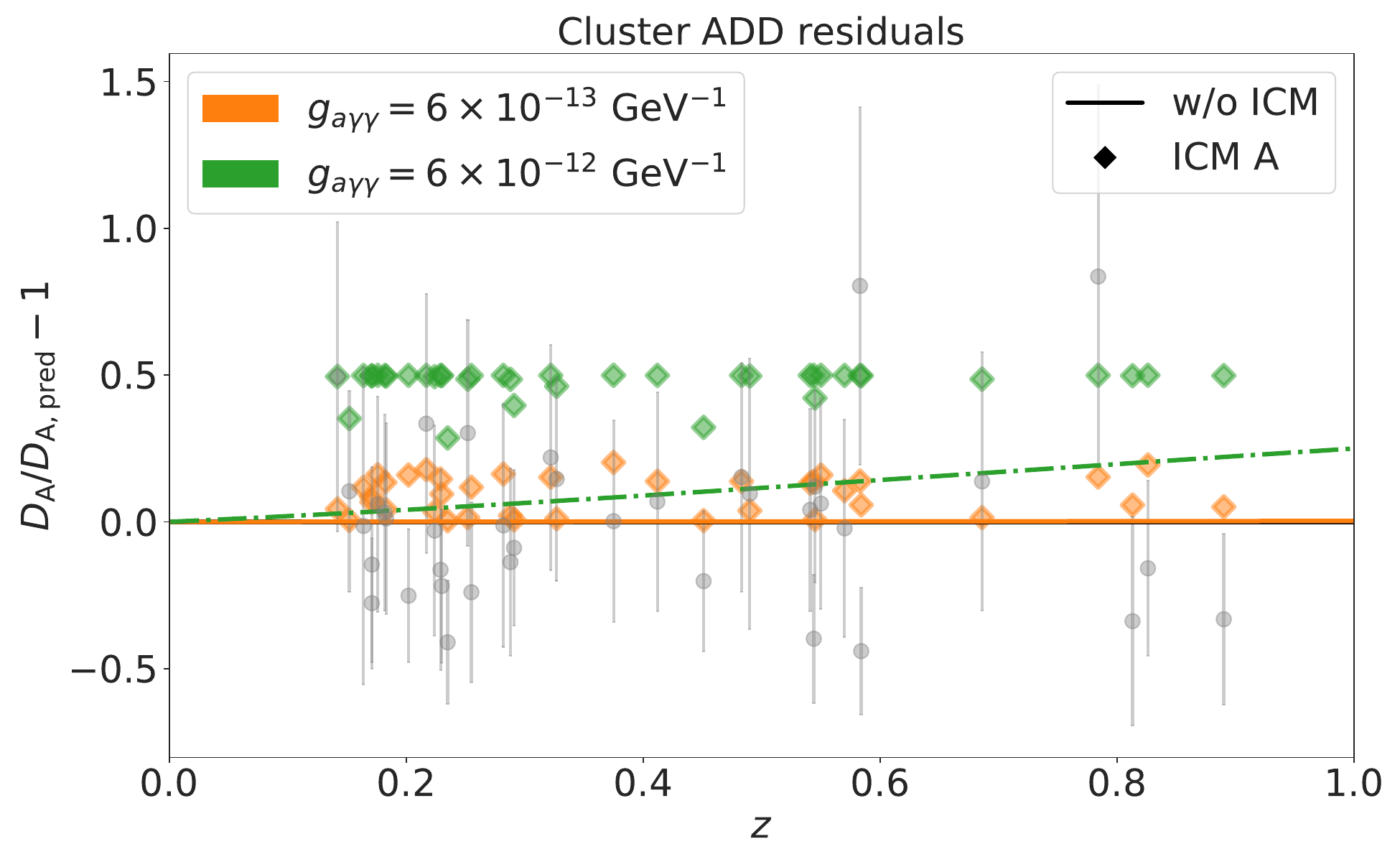}
  \caption{Residuals of the Pantheon SNIa apparent magnitude (left) and the cluster ADD (right) data, compared to a $\LC$ benchmark with $\Omega_\Lambda=0.69$, $H_0=69~\km \; \sec^{-1} \Mpc^{-1}$, and $M=-19.39$. The colors denote the deviation from this benchmark in these observables, for $n_{e,\igm} = 1.6\times 10^{-8}~\cm^{-3}$, $m_a = 10^{-16}~\eV$, and both $\gag = 6 \times 10^{-13} ~ \GeV^{-1}$ (solid, orange) and $\gag = 6 \times 10^{-12} ~ \GeV^{-1}$ (dot-sahed, green). In the right panel, both the case with (diamonds) and without (lines) ICM effects are presented. For the former, we use the magnetic field model A.
  \label{fig:residuals}
}
\end{figure}

To further illustrate this last point, \Fig{fig:residuals} shows the residuals for the Pantheon apparent magnitude (left panel) and cluster ADD (right panel) data, compared to a $\LC$ model with $\Omega_\Lambda=0.69$, $H_0=69~\km \; \sec^{-1} \Mpc^{-1}$, and $M=-19.39$. We also plot the effects of the axion-photon conversion on those observables, for $n_{e,\igm} = 1.6\times 10^{-8}~\cm^{-3}$, $m_a = 10^{-16}~\eV$, and both $\gag = 6 \times 10^{-13} ~ \GeV^{-1}$ (orange) and $\gag = 6 \times 10^{-12} ~ \GeV^{-1}$ (green). Note that in both panels the disagreement due to IGM conversion grows with redshift, as the effect gets stronger for the more distant sources. In the right panel, for the cluster ADD, we consider both cases where we only keep the IGM conversion and ignore ICM effects (lines), and where we include ICM effects with the model A from \Eq{eq:modA} for the ICM magnetic field (diamonds). Note that since each cluster has different parameters for its double-$\beta$ profile, the ADD with ICM effects is different for each cluster. Also note that the presence of ICM conversion is the dominant contribution to the modification of the ADD distances to clusters, overshadowing the $z$-dependent IGM effect behind, and making the bounds independent of $B_\igm$.

\begin{figure}[h]
  \centering
  \includegraphics[width=0.8\textwidth]{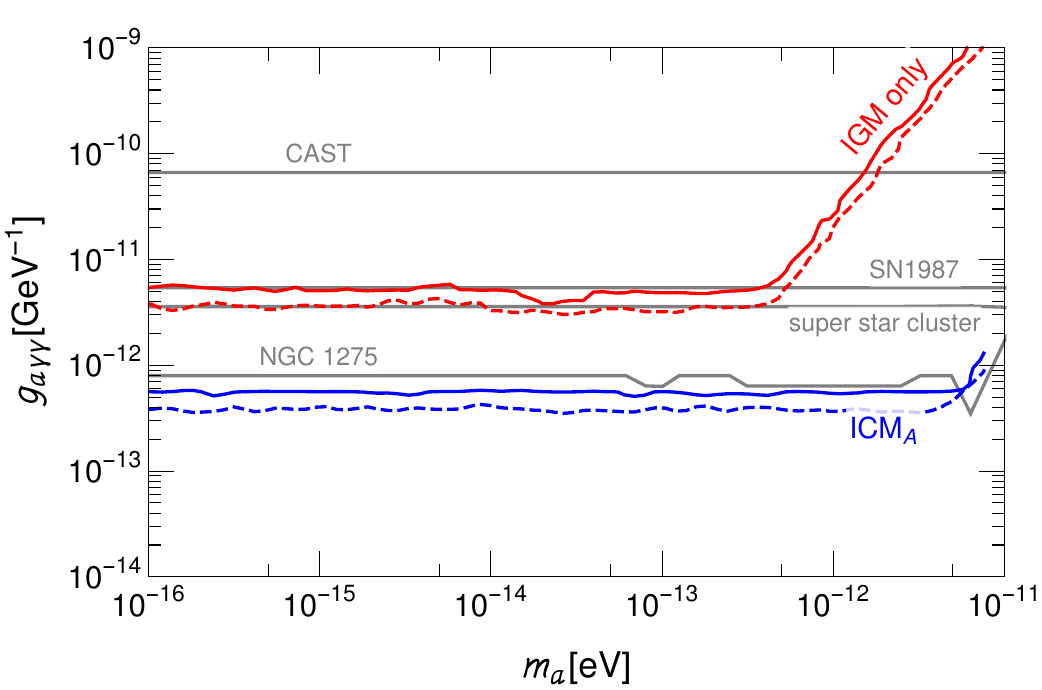}
  \caption{95\% C.L. upper limits on $\gag$ as a function of $m_a$. The solid curves are from $\mL_\late$ while the dashed curves are from $\mL_\early$, assuming $B_\igm=1~\nGa$ and $s_\igm = 1~\Mpc$. To avoid clumsiness, we only show the upper limits from either assuming no ICM conversion effects on the galaxy cluster data (top red curves) or assuming model A in Eq.~\eqref{eq:modA} for the effect (lower blue curves). The upper limits for model B and C in Eqs.~\eqref{eq:modB} and \eqref{eq:modC} are in between them. We also show several existing bounds (grey lines) for comparison: CAST~\cite{Anastassopoulos:2017ftl}; SN1987a~\cite{Payez:2014xsa}; X-ray searches from super star cluster~\cite{Dessert:2020lil} and X-ray spectroscopy from AGN NGC 1275~\cite{Reynolds:2019uqt}.
  \label{fig:photobound}
}
\end{figure}

Lastly, we want to compare our results with existing studies in the literature, which is shown in \Fig{fig:photobound}. For readability, we only show the 95\% C.L. upper limits from either assuming no ICM conversion effects on the galaxy cluster data or assuming model A in \Eq{eq:modA} for the effect. The upper limits for model B and C in \Eqs{eq:modB}{eq:modC} are in between them. In the figure, we also show several other strong bounds on $\gag$ in the same mass range from CAST~\cite{Anastassopoulos:2017ftl}, SN1987a~\cite{Payez:2014xsa} (note that \cite{Bar:2019ifz} proposes a looser bound, due to an alternative modeling of the neutrino emission), X-ray searches from super star cluster~\cite{Dessert:2020lil} and X-ray spectroscopy from AGN NGC 1275~\cite{Reynolds:2019uqt} (note that the ICM magnetic field modeling for NGC 1275 bound is questioned in~\cite{Libanov:2019fzq}). We could see that,
\begin{itemize}
    \item the weakest limit we have, assuming that no X-ray photon-axion conversion in ICM, leads to a bound comparable to existing ones from SN1987a and super star cluster: $\gag \lesssim (4-5) \times 10^{-12}~\GeV^{-1}$ for $m_a \lesssim 5 \times 10^{-13}~\eV$, assuming $B_\igm=1~\nGa$ and $s_\igm = 1~\Mpc$. For other IGM benchmarks, the bounds should be scaled by $\bl( \nGa/B_\igm \br) \bl( \sqrt{\Mpc / s_\igm} \br)$ accordingly for light axions. 
    
    \item if the magnetic field in ICM is described by model A in Eq.~\eqref{eq:modA}, the strongest limit we have pushes $\gag \lesssim (5 -6) \times 10^{-13}~\GeV^{-1}$ for $m_a \lesssim 5 \times 10^{-12}~\eV$. As mentioned above, these bounds are independent of $B_\igm$ and $s_\igm$. Note that to avoid a busy plot, we do not show the bounds assuming model B and C in Eqs.~\eqref{eq:modB} and \eqref{eq:modC}. They are weaker than the one from model A but still stronger than the weakest limit assuming only IGM conversion.
\end{itemize}

Note that axions in the narrow mass range $(6 \times 10^{-13} - 10^{-11})~\eV$ are ruled out by superradiance of stellar black holes~\cite{Arvanitaki:2014wva} and for even lighter axions with mass around or below $10^{-20}$ eV, there exists interesting constraints on $\gag$ from AGN~\cite{Ivanov:2018byi}, protoplanetary disk polarimetry~\cite{Fujita:2018zaj} and CMB birefringence~\cite{Fedderke:2019ajk}, which we do not show in the figure. In addition, the distortion of CMB spectrum due to $\gamma-a$ conversion only places strong bounds at $m_a > 10^{-14}\; \mathrm{eV}$ \cite{Mirizzi:2005ng, Mirizzi:2009nq}, which scales with $B_\igm$.

It has been noted in~\cite{Dror:2020zru} that for ultralight axions, cosmological considerations requiring axions to have a matter-power spectrum that matches that of cold dark matter constrains the magnitude of the axion couplings to the visible sector. As a result, at least part of the parameter space the cosmic distance measurements could probe is associated with non-trivial axion models, in which axions have an abnormally large coupling to photons, as constructed in~\cite{Farina:2016tgd, Agrawal:2017cmd, Agrawal:2018mkd, Dror:2020zru}.

\section{Conclusions}
\label{sec:concl}

In this paper, we show the axion-photon coupling can be strongly constrained by combining several cosmological distance measurements, including luminosity distances to SNIa, angular diameter distances to galaxy clusters, BAO angular size and etc. In contrast to previous practices parametrizing $D_L = D_A(1+z)^{2+\epsilon}$, we demonstrate that the axion-photon oscillation modifies both the luminosity and angular distances in different, non-trivial ways, which cannot be easily captured by a single parameter $\epsilon$. In particular, whether the non-conservation of photon flux affects a measurement is determined by the experimental observable instead of a universal cosmological parameter. This is the reason behind the fact that ADDs from the BAO dataset are not directly affected by the oscillation, whereas those from galaxy cluster datasets could be strongly affected. For the same reason, we avoid using existing results derived from analyses that can be affected by the presence of axion-photon coupling, such as $H_0$ from SH0ES. Instead, we only use the determination of the absolute magnitude $M$ of SNIa from SH0ES.

When axion-photon conversion in ICM is neglected, which serves as a conservative benchmark to avoid the uncertainty of the magnetic field in ICM, we derive a bound comparable to existing bounds from SN1987a and super star clusters. These bounds are effectively constraints on $(\gag \times \frac{B_\igm}{1\;\nGa} \times \sqrt{\frac{s_\igm}{1~\Mpc}})$ for small axion masses, and therefore a direct measurement of $B_\igm$ and $s_\igm$ would fix exactly where this bound lies in the $(m_a, \gag)$ parameter space. On the other hand, the inclusion of X-ray axion conversion in ICM makes the bound even stronger, no matter what ICM magnetic model we choose, and these bounds are entirely independent of the IGM parameters. In particular, model A of the magnetic field in ICM pushes the bound an order of magnitude stronger. A better understanding of the magnetic field in ICM could help reduce the uncertainties associated with its modeling. In addition, positive detection of axion-photon coupling from future experiments probing axion-photon coupling in this mass range \cite{Berlin:2020vrk,Obata:2018vvr,Kahn:2016aff,Liu:2019brz} could help fix these bounds. Lastly, with future improvements in the precision of cosmic distance measurements, a better determination of late-time Hubble diagram $H(z)$ is expected, which could further improve the sensitivity to possible departures from the $\Lambda$CDM prediction due to photon-axion conversion.

\section*{Acknowledgments}
We thank David Pinner for early collaboration of this project. 
We thank Prateek Agrawal, Michael Geller, Dan Hooper, Matt Reece, Martin Schmaltz, Yu-Dai Tsai and Tomer Volansky for discussions at different stages of the project, as well as the anonymous referee, whose suggestions helped improve this paper.
MBA and JF are supported by the DOE grant DE-SC-0010010 and NASA grant 80NSSC18K1010.
CS is supported by the Foreign Postdoctoral Fellowship Program of the Israel Academy of Sciences and Humanities, partly by the European Research Council (ERC) under the EU Horizon 2020 Programme (ERC-CoG-2015 - Proposal n. 682676 LDMThExp), and partly by Israel Science Foundation (Grant No. 1302/19).

\appendix

\section{Brightening Supernovae with axions and the Hubble crisis}
\label{app:A}

In this appendix, we will discuss the interesting possibility of using axions to solve the Hubble crisis between early and late time measurements. This is not directly related to the main goal of our paper but has some similar ingredients, such as photon-axion conversion in IGM due to the magnetic fields. We first discuss some minimum requirements for this possibility and demonstrate why it does not work, at least for some minimal models. Ref.~\cite{Knox:2019rjx} also briefly discusses this possibility and comments on the potential observational challenges it faces, e.g., to explain other late-time datasets such as strong lensing~\cite{Birrer:2020tax}. We will provide a simple argument why this idea could not work even if we simply try to reconcile the SH0ES and Planck results.

The basic idea is that SNIa's further away on the cosmic distance ladder actually appear {\it brighter} than they would be in pure standard $\Lambda$CDM, because they also produce axions, which convert to photons en route and increase the net photon flux observed. Without correcting for the axion effects, the SNIa's further away will appear to be closer to us than they actually are. Thus the deduced $D_L$'s of brightened SNIa's are shorter, resulting in a larger deduced $H_0$, compared to its true value. 
More precisely,  the effective luminosity distance $D_L^\mathrm{eff}$ from the observed flux of photons, $F_\gam^\mathrm{obs}$ in SH0ES is given by
\beq
    D_L^\mathrm{eff} \sim \frac{cz}{H_0^\mathrm{SH0ES}} = \sqrt{\frac{L_\mathrm{SN}}{4\pi F_\gam^\mathrm{obs}}}\ ,
\eeq
where $L_\mathrm{SN}$ is the luminosity of SNIa's. The Hubble value today measured by SH0ES is related to that inferred from Planck data as $H_0^\mathrm{SH0ES} = H_0^\mathrm{Planck} (1+\epsilon)$, where $\epsilon \sim 10\%$. Therefore, assuming $H_0^\mathrm{Planck}$ is the true value of the Hubble rate today and to reconcile the late-time and early-time measurements, we need the observed photon flux to be enhanced by $\sim$20\% compared to the flux without contribution from axions converting into photons. Using the formalism in Sec.~\ref{sec:axion-photon}, we have observed photon intensity from SNIa further away (e.g. at redshift $z\sim 0.1$, or a distance of $y\sim 1$~Gpc away), enhanced by a factor of about 1.2:
\beq
\Pgg (y) =  e^{-x}+ A \bl( 1 - e^{-x} \br) \approx 1.2, \quad {\rm where} \; x= - \frac{1}{s} \int\limits_0^y \dd y' ~ \ln \bl( 1 - \frac{3}{2} P_0(y') \br) >0 . 
\eeq
To satisfy the equation above, we need 
\beq
x \gg 1,  \quad A \equiv \frac{2}{3} \left(1+\frac{I_a^0}{I_\gamma^0}\right) \approx 1.2 \Rightarrow I_a^0 \approx 0.8 I_\gamma^0.
\eeq
Thus we need an initial axion flux $I_a^0$ almost as large as the photon flux $I_\gamma^0$ emitted by SNIa further away to solve the Hubble crisis in this scenario! 

This poses the first challenge to this potential solution. As shown in Ref.~\cite{Grossman:2002by}, the initial axion flux is negligible considering direct axion productions, namely, non-resonant conversions of photons in the SNIa's magnetic fields and in the magnetic fields of their host galaxies. One possibility that was ignored is the resonant conversion of photons to axions. In general, it is not easy to generate a large initial axion flux through resonant conversions, of which the general conditions required could be found in~\cite{PhysRevD.37.2039, PhysRevD.37.1237}. One necessary but not sufficient condition is to have a resonant shell in or near the SN, at which $m_a$ matches the plasma photon mass $m_\gamma$. In a 
SNIa with about one solar mass and a radius of order $10^{15}$ cm (the characteristic radius at 10 days when SNIa reaches its peak luminosity after the explosion of its progenitor white dwarf), the average electron density  corresponds to a plasma photon mass $\sim 10^{-5}$ eV. In the interstellar medium of the host galaxy outside SN, the plasma photon mass is of order $10^{-11}$ eV. Thus, to have resonant conversions inside or near SN, the axion mass has to be $m_a \gtrsim 10^{-11}$ eV.

On the other hand, for axion masses $m_a \gtrsim 10^{-11}$ eV, the photon-axion conversion probability is negligible in IGM. For this axion mass range, the conversion probability in a single magnetic domain is approximately 
\beq
P_0 \approx 2\left(\frac{\gag B \omega}{m_a^2}\right)^2 \approx 10^{-17} \left(\frac{10^{-11}\,{\rm eV}}{m_a}\right)^4 \left(\frac{10^{11}\,{\rm GeV}}{\gag^{-1}}\right)^2 \left(\frac{B}{1\,{\rm nG}}\right)^2  \left(\frac{\omega}{{\rm eV}}\right)^2. 
\label{eq:P0largemass}
\eeq
The probability of axion-photon conversion remains tiny after photons/axions travel over $10^3 - 10^4$ domains from a source Gpc away. This is consistent with our discussion in the main text. We only see a strong bound for $m_a \lesssim 10^{-13}$ eV, in which the photon-axion conversion in IGM becomes non-negligible. 

In summary, to have axions brighten SNIa, we need resonant conversions inside or near SN in order to generate an initial axion flux as large as the initial photon flux. We also need more axions converting into photons in the IGM rather than the other way around. Yet as we show above by considering some simple necessary conditions for the scenario to work, the two requirements mentioned point towards very different axion mass ranges.
 
As bold model builders, we could consider more complicated scenarios, e.g, a photon-dark photon-axion system, similar to the setup in Ref.~\cite{Choi:2018mvk} for a different purpose. Then in the IGM, it is the dark magnetic field, which could be much larger than the ordinary magnetic field, that converts axions into photons or vice versa. Yet even considering a large dark magnetic field of order $10 \mu$G as considered in~\cite{Choi:2018mvk}, we could see from Eq.~\eqref{eq:P0largemass} that the photon-axion conversion probability is still tiny for axion mass above $10^{-11}$ eV. We will leave it for interested readers to explore further whether there are loopholes in our arguments.

\bibliography{ref}
\bibliographystyle{utphys}
\end{document}